\documentclass[twocolumn,english]{revtex4-1}
\usepackage[T1]{fontenc}
\usepackage{float}
\usepackage{units}
\usepackage{amsmath}
\usepackage{amssymb}
\usepackage{graphicx}
\usepackage{esint}
\usepackage{babel}

\usepackage{physics}

\begin{document}

\title{Hight temperature generation of entangled light in parametrically excited micro and nano-ring ressonators for integrated optics}

\author{D. M. N. Magano}
\author{P. A. M. Macedo}
 \affiliation{
 Departamento de F\'{i}sica e Astronomia de Faculdade de Ci\^{e}ncias, Universidade do Porto, Rua do Campo Alegre, 4150-007 Porto, Portugal}
\author{A. Guerreiro}
 \affiliation{
 Departamento de F\'{i}sica e Astronomia de Faculdade de Ci\^{e}ncias, Universidade do Porto, Rua do Campo Alegre, 4150-007 Porto, Portugal}
 \affiliation{
 INESC TEC, Rua do Campo Alegre, 4150-007 Porto, Portugal}

\date{\today}

\begin{abstract}
We propose a physical scheme to generate entangled light at high temperatures through the excitation of vacuum fluctuations of the electromagnetic field using periodic modulations of a refractive index of a ring resonator. We consider the processes of dissipation and decoherence resulting from the coupling of the system with an environment at finite temperature and show that the total amount of entanglement generated does not depend on the temperature, which only affects the time necessary for the system to produce such an amount of entanglement.
\end{abstract}

\maketitle

\textbf{\textit{Introduction}}
The generation of entangled light in integrated nanophotonic systems is of paramount importance to accomplish on chip quantum optical information processing. Entanglement continues to be a crucial resource supporting many quantum information technologies, from quantum computing to quantum security, and its generation is part of the effort to produce quantum sources that can deliver reproducible light with custom specific properties (e.g. low to single photon and squeezed light).

The conventional method of producing entangled light is based on degenerated parametric down conversion (PDC) in nonlinear optical media, such as lithium niobate or aluminum nitrate crystals \cite{guo,saleh,yang,christ}. To improve the efficiency and entanglement yield of this process without excessively increasing the input power there have been put forward several proposals that realize PDC in nanocavity ring resonators (NRR). Indeed, NRR have been exceptionally successful in storing and amplifying light \cite{JOUR}, but delivered relatively moderate results when used to produce entangled light. The reason for this can be found, not only on the low nonlinear coefficients of the materials typically used, but also on the difficulty to combine the geometric resonance conditions of the ring with the adequate phase matching conditions required by PDC, as well as, on the challenge to shield the entangled photons from the destructive effects associated with decoherence as they circle round the ring. Furthermore, the entanglement produced by PDC is embodied in the polarization states of light, which may be difficult to preserve in integrated optics and requires specially designed waveguides that preserve polarization. So far, the production of  light with some amount of entanglement has been shown in NRR with radius as small as $5 \mu m$ \cite{azzini,wakabayashi,silverstone,grassani}.

However, PDC is not the only process that generates entangled light. Another family of effects exists that produce entanglement and are based on parametric amplification (PA) of light via temporal modulation of the optical properties of the medium in which light propagates. Contrary to PDC, the change in the effective refractive index of the medium is not produced by the photons being entangled, but rather by some independent and external source via Kerr effect (or other nonlinear process). In this case, there are no phase matching conditions and the entangled light generated is free to satisfy other resonance conditions, such as those imposed by the closed periodic geometry of a cavity ring. Like PDC, the main difficulty of PA is the demand for large and fast changes of relative refractive index, typically via nonlinear effects.
The recent development of epsilon near zero materials (ENZ) allows to overcome this limitation by yielding variations of refractive index of the order of unity that occur within ultrafast (femtosecond) timescales \cite{Kinsey2015,Shaltout2015,kaipurath2016optically}, without dramatically increasing the loses and the associated dissipation and decoherence processes.

Another noticeable aspect of PA is that entanglement is produced not in the polarization of light, but rather in momentum, since PA entangles photons going in opposite directions. This is a consequence of the spatial-temporal symmetry breaking induced by the modulation of the refractive index, and which favours the separation of the bunches of entangled photons as they exit a ring resonator. If needed, this type of entanglement can be converted into polarization entanglement \cite{Boschi}.

In this work we present a proposal to generate entangled light via PA in NRR and an analyse the optimal conditions for enhanced entangled photon emission even with hard decoherence conditions associated with high temperature operation. This letter is organized as follows: in the next section we present the quantum model for light in NRR and derive the effective Hamiltonian and dynamical equations for the optical modes; the next sections studies the dynamics of Gaussian states of light at both zero and finite temperatures. The last section presents the conclusions.

\begin{figure}
    \centering
    \includegraphics[width=\columnwidth]{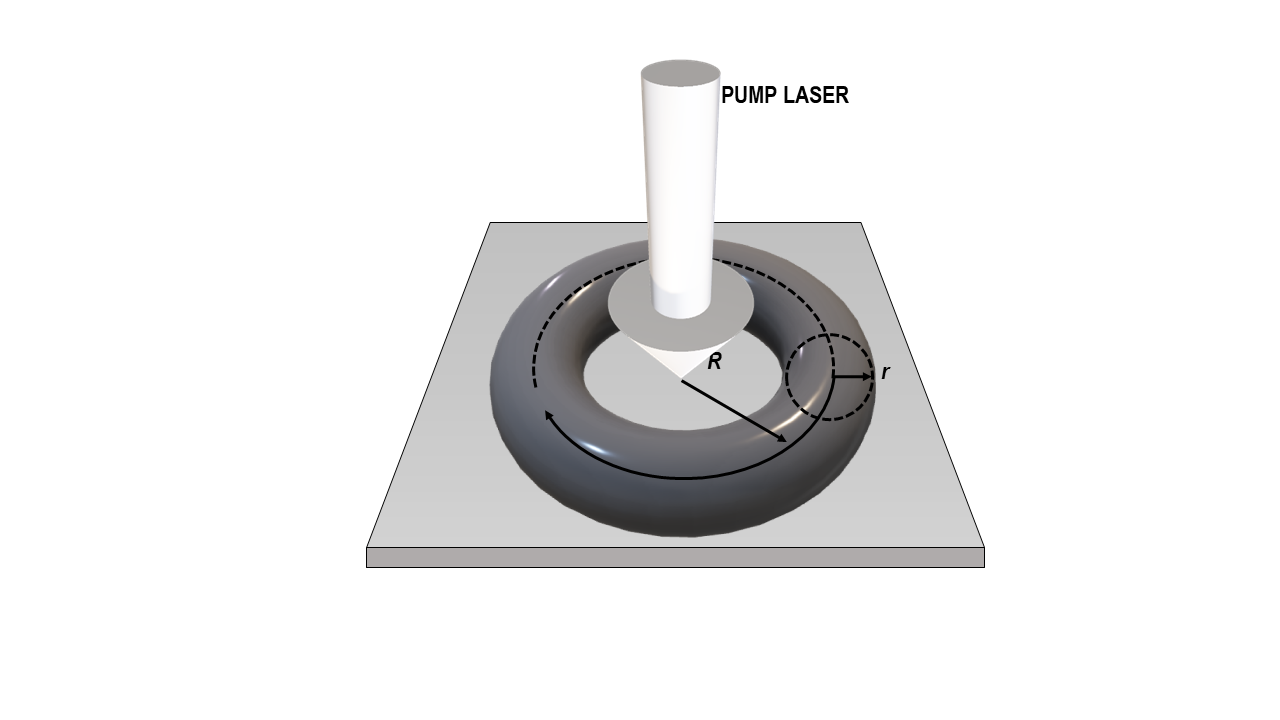}
    \caption{Scheme of the torus cavity constituting the NRR.  constitutes a to}
    \label{Scheme}
\end{figure}

\textbf{\textit{Physical model}}
The NRR is modeled as a toroidal cavity filled with a nonlinear optical material (such as a  ENZ dielectric), whose dielectric permittivity $\epsilon$ can be modulated via an external laser as depicted in Figure \ref{Scheme}, while the magnetic permissivity $\mu$ is assumed to remain constant for simplicity. 
The normal modes $\mathbf{\Phi}_\mathbf{k}$ of the field inside the NRR for $\dot{\epsilon}=0$ are subjected to  boundary conditions on the surface of the torus and are calculated in Supplementary Material A. They constitute an orthonormal basis of the classical Hilbert space associated with the electromagnetic field. Each normal mode satisfies respectively the wave equation  and the closure condition
\begin{gather}
    \laplacian \mathbf{\Phi}_{+\mathbf{k}}=-\frac{1}{\lambda_{\mathbf{k}}^2} \mathbf{\Phi}_{+\mathbf{k}} ,
    \label{property1}\\
     \mathbf{\Phi}_\mathbf{+k}^{*}=\mathbf{\Phi}_\mathbf{-k} . \label{property2}
\end{gather}
and upon quantization, the expansion of the electromagnetic field operators in these modes takes the form
\begin{gather}
\mathbf{\hat{E}} = i \sum_\mathbf{k} \sqrt{\frac{\hslash w_\mathbf{k}(t)}{2 \epsilon(t)}} \left( \hat{a}_\mathbf{k} \mathbf{\Phi}_\mathbf{k}- \hat{a}_\mathbf{k}^{\dagger} \mathbf{\Phi}_\mathbf{k}^{*} \right) \label{expansion1} \\
\mathbf{\hat{B}} = \sum_\mathbf{k} \sqrt{\frac{\hslash}{2 \epsilon(t)  w_\mathbf{k}(t)}} \curl \left( \hat{a}_\mathbf{k}  \mathbf{\Phi}_\mathbf{k} + \hat{a}_\mathbf{k}^{\dagger}  \mathbf{\Phi}_\mathbf{k}^{*} \right), \label{expansion2}
\end{gather}
where the composite index $\mathbf{k}=\{m, \mathbf{n} \}$ summarizes the set of quantum numbers of the normal modes. In  particular the quantum number $m$ characterizes the angular momentum of the mode, while $ \mathbf{n}$ indicates the remaining quantum numbers of the mode. Furthermore, to keep notation simple, it was adopted that $\mathbf{+k} =\{m, \mathbf{n} \}$ and that $ \mathbf{-k}=\{-m, \mathbf{n} \}$, meaning that the composite indexes $\mathbf{k}$ and $\mathbf{-k}$ are related by an inversion of the angular momentum and correspond to light going round in the torus in opposite directions. Also, $w_\mathbf{k}(t)=v(t) / \lambda_\mathbf{k}$, $\hat{a}_\mathbf{k}$ and $\hat{a}_\mathbf{k}^{\dagger}$ are the respectively the time-dependent frequency, the annihilation and creation operators for the normal mode of the field with index $\mathbf{k}$.

The equations of motion for the annihilation and creation operators when $\dot{\epsilon} \neq 0$ are obtained by introducing equations \eqref{expansion1} and \eqref{expansion2} into the Maxwell equations and by taking into account the closure relation \eqref{property2}, thus yielding
\begin{gather}
\dot{\hat{a}}_\mathbf{+k}=-i w \hat{a}_\mathbf{+k} + \frac{\dot{\epsilon}}{4 \epsilon} \hat{a}_\mathbf{-k}^{\dagger} \label{eq_motion1}\\
\dot{\hat{a}}_\mathbf{-k}^{\dagger}=i w \hat{a}_\mathbf{-k}^{\dagger} + \frac{\dot{\epsilon}}{4 \epsilon} \hat{a}_\mathbf{+k} , \label{eq_motion2}
\end{gather}
Equations of motion \eqref{eq_motion1} and \eqref{eq_motion2} can be obtained as the  Heisenberg equations of the effective Hamiltonian
\begin{equation}
    \hat{H} = \hslash f(t)\hat{N} + i \hslash g(t) \hat{S}, \label{hamiltonian}
\end{equation}
with $\hat{N}= \hat{a}_\mathbf{+k}^{\dagger} \hat{a}_\mathbf{+k}+ \hat{a}_\mathbf{-k}^{\dagger} \hat{a}_\mathbf{-k}$, $\hat{S} = \hat{a}_\mathbf{+k}^{\dagger} \hat{a}_\mathbf{-k}^{\dagger}- \hat{a}_\mathbf{+k} \hat{a}_\mathbf{-k}$, $f(t)= w(t)=\frac{\hslash}{\lambda \sqrt{\mu \epsilon(t)}} $, and $g(t)=\frac{1}{4 \epsilon} \frac{d \epsilon}{dt}= -\frac{1}{2} \frac{d \log(f)}{dt}$.
These equations show that the modulation of the dielectric permittivity $\epsilon$ introduced by the external laser couples each mode with $\mathbf{+k}$ only to the mode with $\mathbf{-k}$. In other words, it couples modes with symmetric angular momentum $m$  and corresponding to light propagating in opposite directions inside the NRR. Therefore, from this point and without loss is generality, the analysis will be focused on a generic subspace generated by optical modes with symmetric values of $m$ and identical $\mathbf{n}$.

\textbf{\textit{Dynamics of Gaussian states}}
Under realistic experimental conditions, NRR is in thermal equilibrium with the environment and  the initial photon population corresponds to a thermal state with finite temperature $\mathcal{T}$ \cite{plunien}, which in the limit of zero temperature is the vacuum state.  Both the vacuum and the thermal states are Gaussian states \cite{ferraro} and, since the effective Hamiltonian \eqref{hamiltonian} is a quadratic form in $a_\mathbf{+k}$ and $a_\mathbf{-k}$, the dynamics of the system preserves the Gaussian character of the state \cite{schumaker}. 
In other words, the time evolution of the system maps Gaussian states into Gaussian states and the dynamics of the photon population can be investigated using the continuous variables formalism. In particular, the Wigner function of a Gaussian state has the form
\begin{equation}
    \mathcal{W}(\mathbf{\hat{x}}) = \frac{1}{2\pi \sqrt{\text{det}\mathbf{\Sigma}}} 
    e^{-\frac{1}{2} (\mathbf{\hat{x}}-\overline{\mathbf{x}})^T \mathbf{\Sigma}^{-1} (\mathbf{\hat{x}}-\overline{\mathbf{x}})},
\end{equation}
where $\mathbf{\hat{x}}=[\hat{x}_\mathbf{+k},\hat{p}_\mathbf{+k},\hat{x}_\mathbf{-k},\hat{p}_\mathbf{-k}]^T$ and $\mathbf{\Sigma}=\overline{\mathbf{x}\mathbf{x}^T}-\overline{\mathbf{x}}\overline{\mathbf{x}^T}$ is the covariance matrix. Therefore, a Gaussian state is fully determined by $\overline{\mathbf{x}}(t)$ and $\mathbf{\Sigma}(t)$.

The model for the processes of loss and decoherence resulting from the coupling of the NRR to the environment assumes that the later corresponds to  a Markovian bath at finite temperature $\mathcal{T}$ and is described using a master equation in the Limblad form (similar to \cite{dodonov} and \cite{jakub})
\begin{align}
\dot{\hat{\rho}} = & -i \left[\hat{H},\hat{\rho} \right]
    + \frac{\gamma}{2} \sum_{i j=\pm\mathbf{k}} \Big[
    (\overline{n}+1) \hat{a}_i \hat{\rho} \hat{a}_j^{\dagger} \nonumber\\
  &- (\overline{n}+1/2)(\hat{a}_i^{\dagger} \hat{a}_j \hat{\rho} + \hat{\rho} \hat{a}_i^{\dagger} \hat{a}_j)
    + \overline{n}(\hat{a}_i^{\dagger} \hat{\rho} \hat{a}_j - \hat{\rho}) \Big],\label{Limblad} 
\end{align}
where $\gamma$ is the coupling parameter between the system and the bath, $\overline{n}=[exp(\hbar w_\mathbf{k} / K_B \mathcal{T} ]-1)^{-1}$, is the mean number of quanta in the bath and  $w_\mathbf{k}$ is the frequency of the mode, respectively.
Under these conditions, equation \eqref{Limblad} expressed in terms of $\overline{\mathbf{x}}(t)$ and $\mathbf{\Sigma}(t)$ is
\begin{gather}
    \dot{\overline{\mathbf{x}}}=-\left(\mathbf{M}^T+\frac{\gamma}{2} \mathbf{\Lambda} \right) \overline{\mathbf{x}} \label{evXT} \\
    \dot{\mathbf{\Sigma}}=-\mathbf{M}^T \mathbf{\Sigma} - \mathbf{\Sigma} \mathbf{M} -\frac{\gamma}{2} \left( \mathbf{\Lambda} \mathbf{\Sigma}+ \mathbf{\Sigma} \mathbf{\Lambda} -2 \mathbf{\Lambda} \mathbf{\Sigma}_{\infty} \right), \label{evSigmaT}
\end{gather}
where $\mathbf{M}$, $\mathbf{\Lambda}$ and $\mathbf{\Sigma}_{\infty}$ are defined explicitly in the Supplementary Material B. It should be noticed that $\mathbf{\Sigma}_{\infty}$ is the covariance  matrix for the stationary solution of equation \eqref{evSigmaT} in the absence of a modulation in $\epsilon$, corresponding to the thermal state in equilibrium with the environment, which is assumed to be the initial state of the system before the external laser is switched on.

Applying the unitary transformation $\mathbf{x}'=\mathbf{\Gamma} \mathbf{x}$, with $\mathbf{\Gamma} \equiv  (1/\sqrt{2}) \mathbf{\Lambda}$, introduces an interaction picture where the original two normal modes are decoupled and equations \eqref{evXT} and \eqref{evSigmaT} becomes 
\begin{gather}
    \dot{\overline{\mathbf{x}}}^\prime=-\left(\mathbf{M}^{\prime T}+\frac{\gamma}{2} \mathbf{\Lambda}^\prime \right) \overline{\mathbf{x}}^\prime \label{evXT2} \\
        \dot{\mathbf{\Sigma}^\prime}=-\mathbf{M}^{\prime T} \mathbf{\Sigma}^\prime - \mathbf{\Sigma}^\prime \mathbf{M}^\prime  -\frac{\gamma}{2} \left( \mathbf{\Lambda}^\prime \mathbf{\Sigma}^\prime+ \mathbf{\Sigma}^\prime \mathbf{\Lambda}^\prime -2 \mathbf{\Lambda}^\prime \mathbf{\Sigma}_{\infty} \right), \label{evSigmaT2}
\end{gather}
with $\mathbf{\Lambda}^{\prime } =  \mathbf{\Gamma}\mathbf{\Lambda} \mathbf{\Gamma}^{-1}$ and \begin{equation}
\mathbf{M}^{\prime}(t)=   \mathbf{\Gamma}\mathbf{M} \mathbf{\Gamma}^{-1} =
\begin{bmatrix}
\mathcal{M}_{+}(t) & 0 \\
0  & \mathcal{M}_{-}(t) 
\end{bmatrix},
\label{matrizM}
\end{equation}
corresponding to a time dependent block diagonal matrix composed by two $2\times 2$ matrices $\mathcal{M}_{\pm}(t) \equiv i f(t) \mathcal{\sigma}_y \mp g(t) \mathcal{\sigma}_z$, with $\mathcal{\sigma}_y$ and $\mathcal{\sigma}_z$ being  the Pauli matrices. For clarity of notation, we represent $2\times 2$ matrices using calligraphic letters. 
Since  $\mathbf{M}^{\prime}$ is bloch diagonal in this representation, the system has been decomposed into two independent or decoupled subspaces. As it will be demonstrated in what follows, these subspaces play a critical role in the generation and conservation of entanglement by the system. 

\textbf{\textit{Zero temperature dynamics}}
To establish a reference for the dynamics of the system, we start by considering a system decoupled from the environment at zero temperature, for which the initial state corresponds to the vacuum state of the field and $\gamma = 0$. Then, it is possible to write the solutions of \eqref{evXT2} and \eqref{evSigmaT2} in terms of the evolution operator $\mathbf{U}(t)$ as
\begin{gather}
    \overline{\mathbf{x}}^\prime(t)=\mathbf{U}(t) \overline{\mathbf{x}}^\prime_0 \label{evXTprime}\\
    \mathbf{\Sigma}^\prime(t)= \mathbf{U}(t) \mathbf{\Sigma}_0 \mathbf{U}^T(t),
\end{gather}
where $\mathbf{U}$ is of the form
\begin{equation}
\mathbf{U} =
\begin{bmatrix}
\mathcal{U}_{+} & 0 \\
0  & \mathcal{U}_{-} .
\end{bmatrix}.\label{Udef}
\end{equation}
Replacing \eqref{Udef} into \eqref{evXTprime}  yields the equation for $\mathcal{U}_\pm$
\begin{equation}
    \dot{\mathcal{U}}_\pm (t)= \mathcal{M}_\pm(t) \mathcal{U}_\pm(t) \label{Udyn}
\end{equation}
Using the techniques described in the Supplementary Material B, it is possible to define a set of transformations $\mathcal{S}_\pm$ such that $\dot{\mathcal{U}}_{\pm} = \mathcal{S}_\pm^{-1} \mathcal{U}_h \mathcal{S}_\pm $ and recast \eqref{Udyn} as
\begin{equation}
    \dot{\mathcal{U}_h} = \begin{bmatrix}
0 & 1 \\
-f^2  & 0 
\end{bmatrix}  \mathcal{U}_h .\label{Udyn2}
\end{equation}
This corresponds to the matrix form of the celebrated Hill equation \cite{magnus}. Indeed, if we define 
\begin{equation}
\begin{bmatrix}
\pi(t)\\
y(t)
\end{bmatrix} = \mathcal{U}_h (t) \begin{bmatrix}
\pi_0\\
y_0
\end{bmatrix},
\end{equation}
then, $y$ satisfies the evolution of a parametrically excited oscillator in the absence of losses
\begin{equation}
\ddot{y}+f(t)^2y=0. \label{Hill}
\end{equation}
The Floquet theorem \cite{magnus} ensures that the general solution of this equations is of the form
\begin{equation}
    y(t)=a e^{\nu t} p_1(t) + b e^{- \nu t} p_2 (t),
\end{equation}
where $p_i$ are tow independent periodic functions with the same period as $f$.
The Lyapunov exponent $\nu$ determines the stability of the solution. 
Indeed, if $\nu$ has a real part, then the solution is unstable and grows exponentially in time, corresponding to the case of interest for a sustained generation of photons inside the NRR. The photon yield for times $t>0$ is determined by the Lyapunov exponent, which can be computed using the property \cite{magnus}
\begin{equation}
    \text{Tr}(\Lambda(T))=2 \text{cosh} \left(\nu T \right),  \label{property}
\end{equation}
where $T$ is the period of the modulation of the dielectric permittivity $\epsilon$, as described by $f$. 

To illustrate the dynamics of the photon generation resulting from the parametric excitation, we consider two limiting cases: (i) a sharp modulation of the dielectric permittivity $\epsilon$, typically produced by an ultra-short pulsed external laser; and (ii) a sinusoidal modulation of $\epsilon$, typically produced by the beating of two continuous external laser beams with similar wavelengths.

In the first case,  $f$ is a approximately a step-wise rectangular modulation
\begin{equation}
f(t)=
    \begin{cases}
      f_1, \quad 0< \text{mod}(t/T)<t_1 \\  
      f_2, \quad t_1< \text{mod}(t/T)<t_1+t_2=T  
    \end{cases}
\end{equation}
then \eqref{Hill} reduces to the Meissner equation. For a fixed ratio $f_r =f_2/f_1$, $\nu$ is maximum for $t_i f_i = \pi/2$, such that
\begin{equation}
    \nu=\frac{1}{T} \abs{\log f_r}.
    \label{RessRec}
\end{equation}
On the other hand, if $\epsilon$ undergoes a small sinusoidal perturbation with period $T$, then using perturbative techniques up to the first order, one obtains that
\begin{equation}
f(t)^2=f_0^2 \left[ 1- \frac{\delta \epsilon}{\epsilon} \text{sin} (\Omega t) \right],
\end{equation}
in which case \eqref{Hill} is called the Mathieu equation. This equation supports parametric resonances at specific values of $T$ and excitation amplitudes $\delta \epsilon/\epsilon$, for which the Lyapunov exponent has a real part and the number of photons gets amplified inside the cavity.
The first resonance of this system occurs for $T=\pi/f_0$, for which the Lyapunov exponent is given by
\begin{equation}
    \nu=\frac{f_0}{4} \frac{\delta \epsilon}{\epsilon}.
    \label{RessSin}
\end{equation}

Unlike the phase matching conditions necessary for PDC to occur, which result from a careful balance between the optical properties of the medium and the geometry of the cavity, the parametric resonances associated with PA can be tuned to by adjusting the parameters of the external laser. This provides an essential mechanism for control and optimization of the photon yield of the MNRR.

Unfortunately, the operation of the MNRR at zero temperature is neither achievable in practice nor desirable for real applications, where ideally the system should provide entangled light at room temperature.  Intuitively and in a classical model, the coupling between the MNRR and the environment introduces loss mechanisms. More specifically, the inclusion of strong losses in the Hill equation for classical models can prevent the existence of parametric resonances \cite{nayfeh}. In a quantum model, besides the loss mechanism, the coupling with the environment introduces decoherence mechanisms that in principle injure the production of entangled light. In the next sections we investigate how the coupling to the environment affects not only the photon yield but more importantly, the generation of entangled light.

\textbf{\textit{Finite temperature dynamics}}
The solution at a finite temperature $\mathcal{T}$ is now built upon the result of the previous section. In particular, since $\left[ \mathbf{\Lambda}^\prime,\mathbf{U} \right]=0$, the solution of \eqref{evXT2} is simply
\begin{equation}
    \overline{\mathbf{x}}^\prime(t)= \mathbf{U}_{th} (t) \overline{\mathbf{x}}^\prime_0,
\end{equation}
where $\mathbf{U}_{th}(t)= exp(-\gamma \mathbf{\Lambda}^\prime t ) \mathbf{U}(t)$.  The solution of \eqref{evSigmaT} can be decomposed into the sum of a homogeneous part $\Sigma_h^\prime$ and a particular solution $\Sigma_p^\prime$, say
\begin{equation}
    \Sigma^\prime = \Sigma_h^\prime + \Sigma_p^\prime
\end{equation}
with $\mathbf{\Sigma}_h^\prime (t) = \mathbf{U}_{th} (t)\mathbf{\Sigma}_0 \mathbf{U}_{th}^T(t)$, $\mathbf{\Sigma}_p^\prime (t) = \mathbf{V} (t) \mathbf{\Sigma}_{\infty}$, and 
\begin{gather}
\mathbf{V}(t) = \begin{bmatrix}
\mathcal{V}(t) && 0\\
0 && 0
\end{bmatrix}\\
\mathcal{V}(t) = 2 \gamma  \mathcal{U}_{+} \left[ \int_0^t  (\mathcal{U}_{+}^T \mathcal{U}_{+})^{-1} dt \right] \mathcal{U}_{+}^T .
\end{gather}

These results indicate that the coupling to a Markovian bath at a finite temperature $\mathcal{T}$ essentially amounts to computing $\mathcal{V}$, which means that, for a given shape of $f(t)$, the resonant modulation is independent of temperature and thus the results provided by equations \eqref{RessRec} and \eqref{RessSin} remain valid.

Remarkably, these results do not follow the intuition provided by the classical models of the Hill equation. A close analysis of the matrices $\mathbf{M}^{\prime}$ and $\mathbf{\Lambda}^{\prime}$  shows that in the new variables  $\mathbf{x}'=\mathbf{\Gamma} \mathbf{x}$ the system supports a decoherence free subspace that is decoupled from the environment and isolated from both losses and decoherence, but which still undergoes parametric amplification by the modulation introduced by the external laser. As it shall be shown in the following section, this aspect of the system dynamics not only provides a strong photon yield, but more importantly supports a robust generation of strongly entangled light at any finite temperature.

\begin{figure}
    \centering
    \includegraphics[width=\columnwidth]{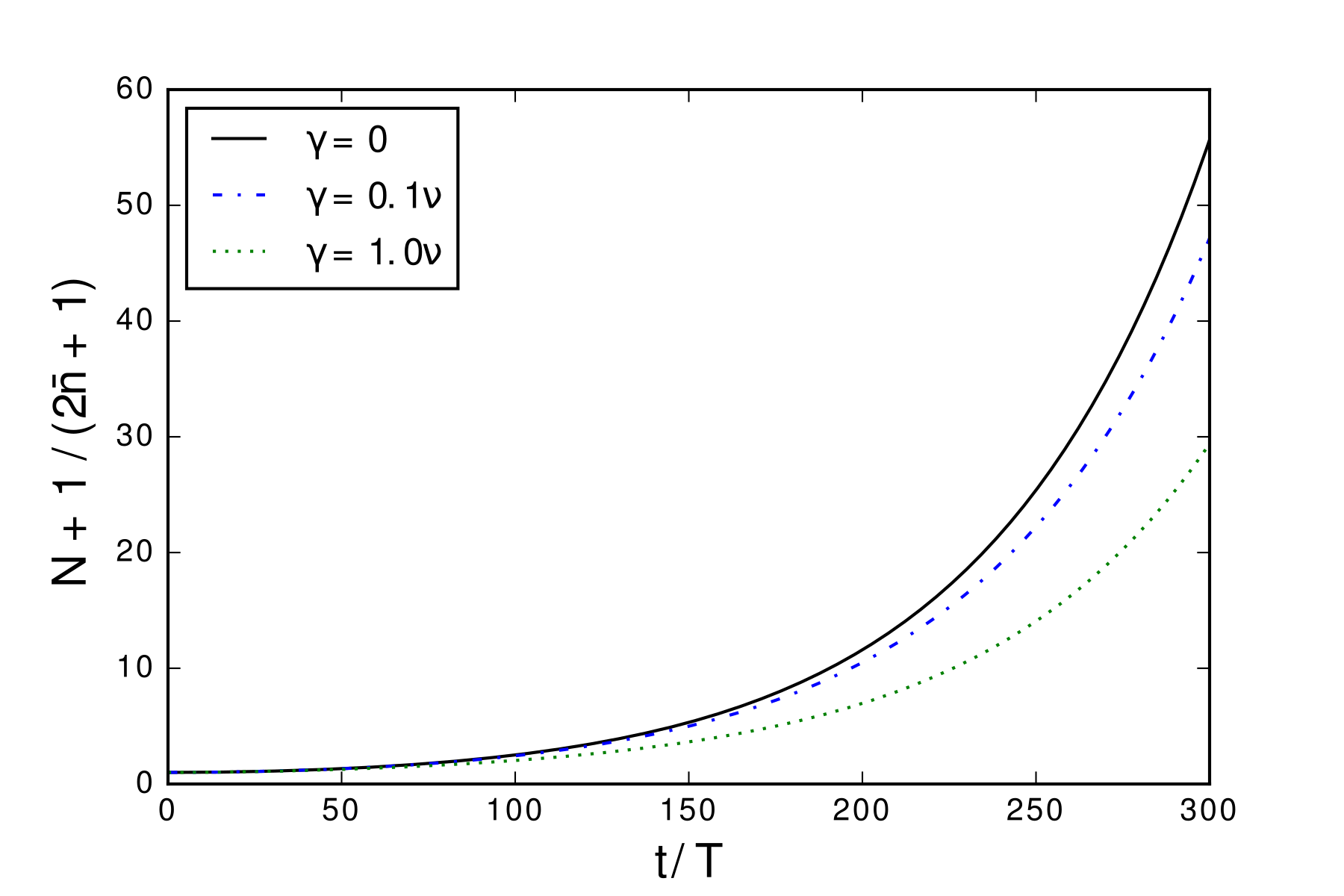}
    \caption{Emission of photons (normalised by $2 \bar{n}+1$) over time for different values of $\gamma$ for a sinusoidal modulation. We considered $\delta \epsilon /\epsilon =0.01$ and used time units in which $T \equiv 1$.}
    \label{PhotonProduction}
\end{figure}

\begin{figure}
    \centering
    \includegraphics[width=\columnwidth]{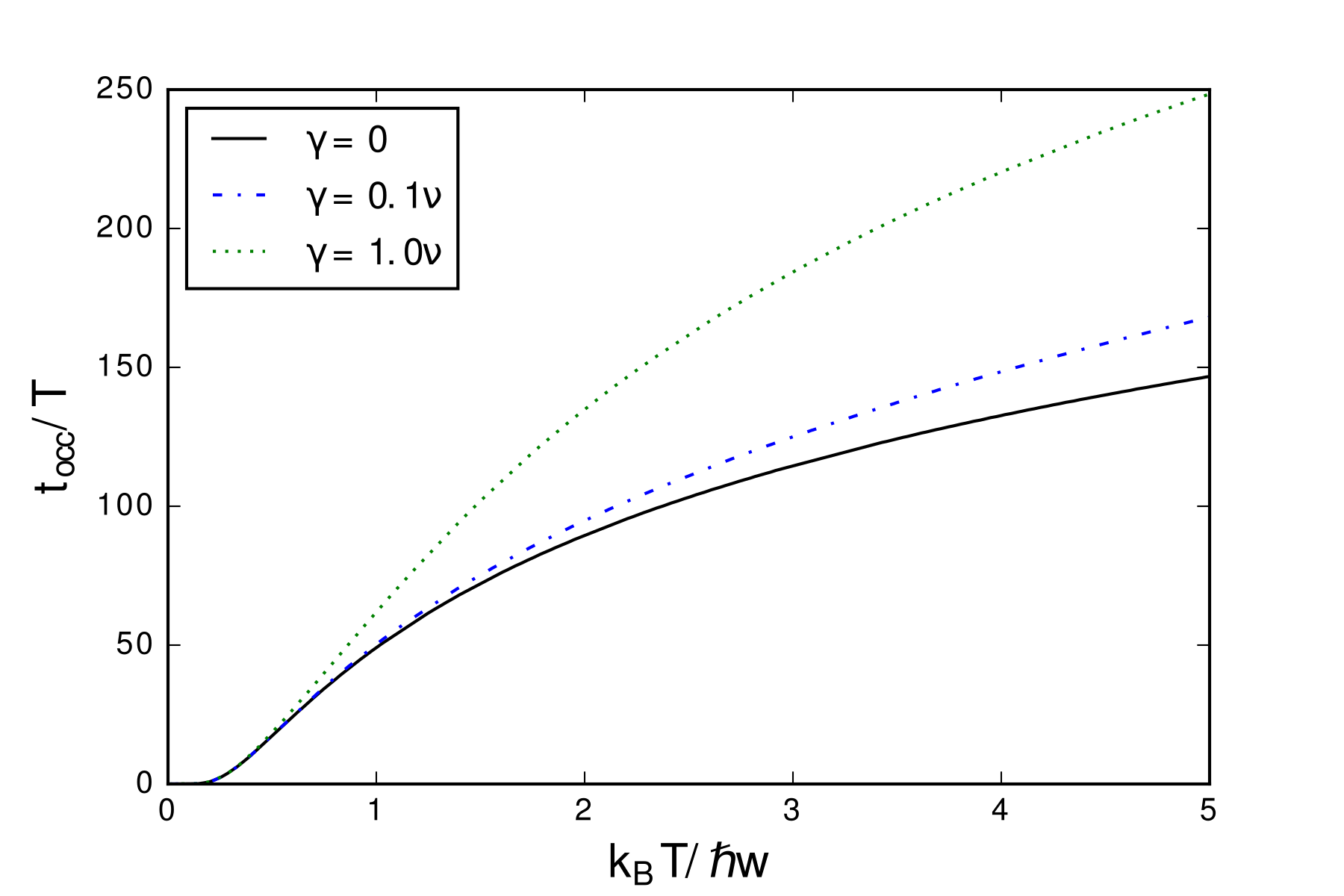}
    \caption{Occurrence time (the time when $E_N$ becomes non-zero) as a function of temperature for different values of $\gamma$ for a sinusoidal modulation. Again, we considered $\delta \epsilon /\epsilon =0.01$ and used the same time units.}
    \label{OccurrenceTime}
\end{figure}

\begin{figure}
    \centering
    \includegraphics[width=\columnwidth]{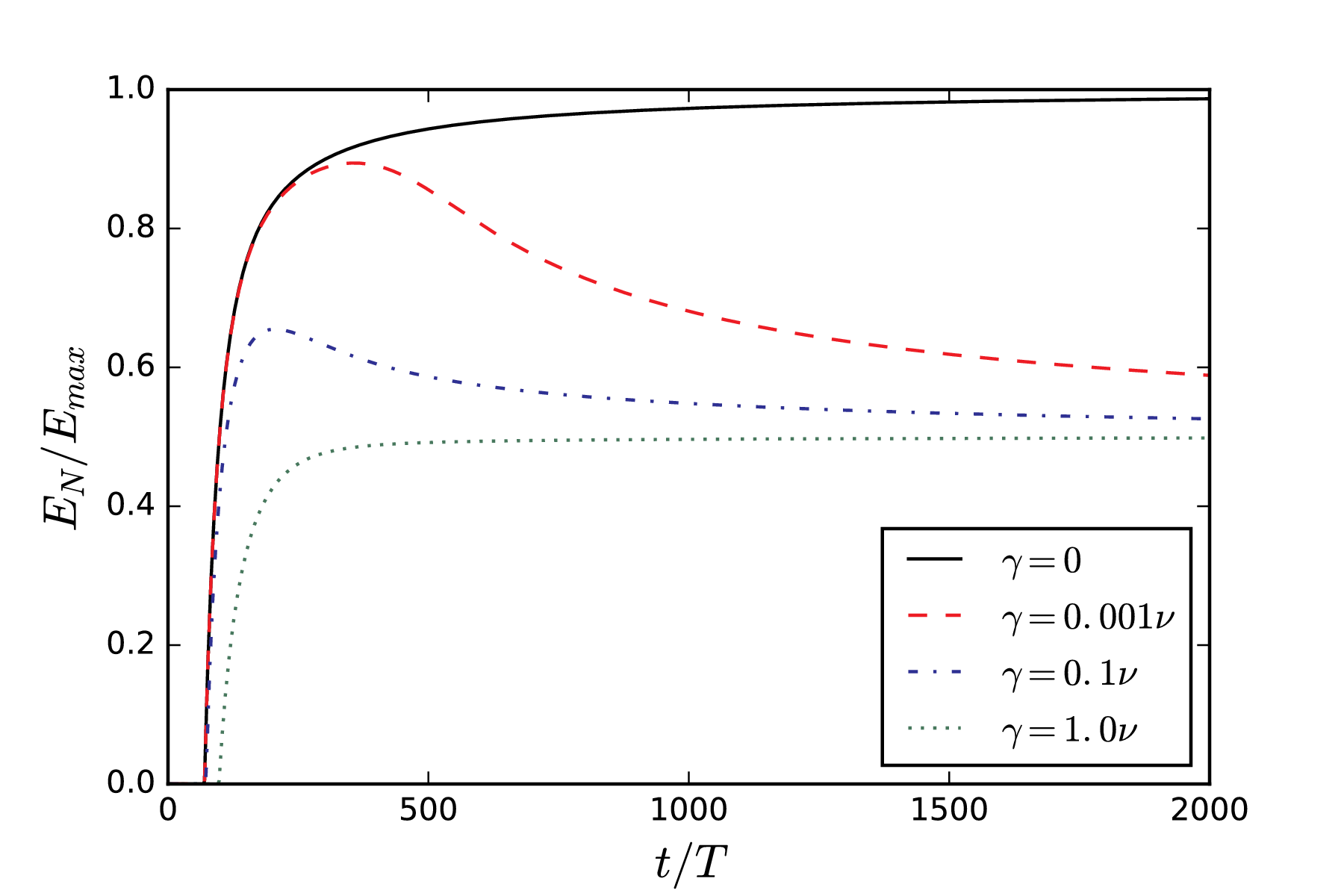}
    \caption{Evolution of the ratio between the logarithmic negativity and the logarithmic negativity of a maximally entangled state in $\mathcal{E}(N)$ for an initial thermal state with $\overline{n}=1$ and for different values of $\gamma$ for a sinusoidal modulation. According to \eqref{Emax}, $E_{max}=\log_2 \expval{N+1}$. Of course, $N$ evolves in time as \eqref{N(t)}. Once again, we considered $\delta \epsilon /\epsilon =0.01$ and used $T \equiv 1$.}
    \label{MaxEntanglement}
\end{figure}

\textbf{\textit{Photon yield and entanglement generation}}
Starting with and initial thermal state and considering a rectangular modulation, the number of photons generated in each mode for $t=n T$ (with $n$ integer) is
\begin{align}
\expval{N+1} = & \frac{2\overline{n}+1}{4} \Big[
    2 \cosh (2 \nu t) +  e^{- \eta_- t} + e^{- \eta_+ t} \nonumber\\
    &+ \left(1-e^{- \eta_{-} t } \right) F_{-}^r+ \left(1-e^{- \eta_{+} t} \right) F_{+}^r \Big] 
\label{N(t)}
\end{align}
while the amount of entanglement between the two modes measured in terms of logarithmic negativity is
\begin{align}
E_N =  \text{max} \Big\{ 0,{} &  
\frac{\nu t}{\log 2}  -\frac{1}{2} \log_2 \left[e^{- t \eta_+}(1-F_+^r) + F_+^r\right]  \nonumber\\
& -\log_2(2 \bar{n}+1)\Big\}
\label{entanglement}
\end{align}
with
\begin{gather}
    F_{\pm}^r=\frac{
     1-e^{2 \gamma t_1}+e^{2 \gamma t_1}\left(1-e^{2 \gamma t_2} \right) f_r^{\pm1} }{1-e^{ \eta_{\pm}T}}\\
     \eta_{\pm}=2(\gamma \pm \nu).
\end{gather}
For the sinusoidal perturbation we get the same results, but now with $F_{\pm}^r$ calculated according to
\begin{equation}
    F_{\pm}^s= \frac{2 \gamma}{\eta_{\mp}} \frac{  1-e^{ \eta_{\mp} T} }{1-e^{ \eta_{\pm}T}}.
\end{equation}
Both cases lead to similar expressions for the average number of photons in the mode and the logarithmic negativity of the state, strongly suggesting that they depend weakly on the particular shape of the modulation. Indeed, in the asymptotic limit $t \rightarrow \infty$, the number of emitted photons becomes
\begin{equation}
\expval{N+1} \sim  \frac{2\overline{n}+1}{4} \left[
     e^{2 \nu t} +  e^{- \eta_- t} (1-F_-) \right] ,
\end{equation}
where $F_-$ can refer to either modulation scenario. Hence, the rate of photon production tends to $2 \nu$, as shown in Figure \ref{PhotonProduction}.
Equation \eqref{entanglement} shows both modulations will always produce entanglement and that the entanglement is proportional to the number of photons produced, as expected. Figure \ref{MaxEntanglement} shows the evolution of the amount of entanglement  as a fraction of a maximally entangled symmetric Gaussian state with the same number of photons $E_{max}(N)$  (see Supplementary Material C)
\begin{equation}
    E_N \leq E_{max}(N)=\log_2 \expval{N+1}.
\label{Emax}
\end{equation}
The results show that, for a system with no coupling to the environment, the  modulation will eventually converge into a maximally entangled state. However, the behavior is different if the system is in contact with a thermal bath. Then, we find that the logarithmic negativity always tends to $E_{max}/2$, independently of the value of $\gamma$. This can be explained by the existence of two independent subspaces in the system: (i) the first is coupled to the environment and undergoes decoherence, so any entanglement generated in the initial times is eventually destroyed for later times; and (ii) the second which is a decoherence free subspace, where the generated entanglement can be stored and accumulated  indefinitely for any arbitrary temperature.
Although entangled light is always generated, the occurrence time necessary for the first pair photons to be produced is delayed as the temperature or the coupling with the bath increase, as shown in Figure \ref{OccurrenceTime}.

To estimate the photon yield and entanglement production, we considered experimental values for this type of systems. For example, at room temperature and for $\lambda \sim 2 \pi R$ it results that  $k_B \mathcal{T} / \hslash w \sim 10^6 R n$, where $n$ is the refractive index of the NRR. For $R$ in nano-scale this implies $k_B \mathcal{T} / \hslash w \sim 10^{-3} n$ while for $R$ in micro-scale it it implies $k_B \mathcal{T} / \hslash w \sim n$. Also, for refractive index characteristic of ENZ materials $n \sim 0.1$ the occurrence time is very small, negligible in fact. These estimates strongly suggest that this type of systems are experimentally viable as micro and (eventually nanoscale) sources of entangled light.

\textbf{\textit{Conclusions}}
In this work we proposed an experimental setting which is capable of supporting and amplifying entanglement at an arbitrary finite temperature. 
In more detail, we considered the excitation of vacuum fluctuations of the electromagnetic field through the periodic modulation of the refractive index of a ring resonator.
We analyzed the optimal conditions for resonance and found that is always possible to amplify radiation, regardless of the loss regime. Furthermore, we found that the tuning between the frequency and the amplitude of the external modulation is the preponderant factor in order to achieve resonance, and not the shape of the modulation.

We were able to verify that the existence of a decoherence free subspace that allows quantum correlations to survive, making entanglement viable even at high temperatures. We also discussed the asymptotic limits for the rate of photon production and for the persistence of entanglement.

Hopefully, in exploring the theory of generation of entangled light, these results will add a contribution to the rising of quantum information technologies.

\include{suplementary}

\bibliographystyle{apsrev4-1}

\end{document}